\documentclass[aps,pra,preprintnumbers,showpacs,tightenlines]{revtex4}
\usepackage{amssymb}
\usepackage{amsmath}
\usepackage{graphicx}
\usepackage{epsfig}
\usepackage{subfigure}
\usepackage{amsfonts}
\usepackage{CJK}

\begin{document}

\title{Generation of GHZ entangled states of photons in multiple cavities
via a superconducting qutrit or an atom through resonant interaction}

\author{Chui-Ping Yang$^{1,2}$, Qi-Ping Su$^{1}$, and Siyuan Han$^{3}$}

\address{$^1$Department of Physics, Hangzhou Normal University,
Hangzhou, Zhejiang 310036, China}

\address{$^2$State Key Laboratory of Precision Spectroscopy, Department of Physics,
East China Normal University, Shanghai 200062, China}

\address{$^3$Department of Physics and Astronomy, University of
Kansas, Lawrence, Kansas 66045, USA}

\date{\today}

\begin{abstract}
We propose an efficient \textrm{method} to generate a GHZ entangled state of
$n$ photons in $n$ microwave cavities (or resonators) via resonant
interaction to a single superconducting qutrit. The deployment of a qutrit,
instead of a qubit, as the coupler enables us to use resonant interactions
exclusively for all qutrit-cavity and qutrit-pulse operations. This unique
approach significantly shortens the time of operation which is advantageous
to reducing the adverse effects of qutrit decoherence and cavity decay on
fidelity of the protocol. Furthermore, the protocol involves no measurement
on either the state of qutrit or cavity photons. We also show that the
protocol can be generalized to other systems by replacing the
superconducting qutrit coupler \textrm{with }different types of physical
qutrit, such as an atom in the case of cavity QED, to accomplish the same
task.
\end{abstract}

 \maketitle
\date{\today}

\begin{center}
\textbf{I. INTRODUCTION}
\end{center}

Entanglement is one of the most fascinating features of quantum mechanics
and plays an important role in quantum communication and quantum information
processing (QIP). During the past decade, experimental preparation of
entanglement with eight photons via \textit{linear optical devices} [1],
eight ions [2], three spins [3], two atoms in microwave cavity QED [4], two
atoms plus one cavity mode [5], or two excitons in a single quantum dot [6]
has been reported.

Over the past ten years, there has been much interest in quantum information
processing with superconducting qubits. By having qubits coupled through
capacitors, entangling two [7] or three superconducting qubits [8] has been
experimentally demonstrated. In addition, a tripartite entanglement
consisting of a superconducting qubit and two microscopic two-level systems
has been reported recently [9].

On the other hand, physical systems composed of cavities and superconducting
qubits such as transmon and phase qubits are considered as one of the most
promising candidates for quantum information processing. For the sake of
simplicity, hereafter the term cavity refers to either a three-dimensional
cavity or any other types of resonant structure such as a coplanar waveguide
(CPW) resonator, a microstrip resonator, or even a lumped circuit LC
resonator. In this circuit QED approach, a cavity acts as a quantum bus
which can mediate long-distance, fast interaction between distant
superconducting qubits [10-14]. Theoretically, it was predicted earlier that
the strong coupling limit can readily be achieved with superconducting flux
qubits [15] or charge qubits [12] coupled to resonant cavities, which has
been experimentally demonstrated soon after [16,17]. Based on circuit QED, a
large number of theoretical schemes for creating entangled states with
superconducting qubits in single cavities have been proposed [10,15,18-25].
In addition, various two-qubit or three-qubit entangled states have been
experimentally demonstrated with superconducting qubits coupled to single
cavities [26-30]. All of these theoretical and experimental works are
focused primarily on entanglement of superconducting qubits coupled to a
single cavity, which has paved the way for fundamental tests of quantum
entanglement and made superconducting qubit circuit QED very attractive for
quantum information processing.

Recently, attention has been progressed to entanglement generation of qubits
or photons resided in multiple cavities because of its importance to
scalable QIP. Within circuit QED, several theoretical proposals for
generation of entangled photon Fock states of \textit{two} resonators have
been presented [31,32]. Moreover, by using a superconducting phase qubit
coupled to two resonators, recent experimental demonstration of an entangled
NOON state of photons in \textit{two} superconducting microwave resonators
has been reported [33].

In this paper, we focus on the preparation of GHZ
(Greenberger-Horne-Zeilinger) entangled states of photons in \textit{multiple%
} cavities. The GHZ entangled states are of great interest to the
foundations of quantum mechanics and measurement theory, and are an
important resource for quantum information processing [34], quantum
communication (e.g., cryptography) [35-37], error correction protocols [38],
and high-precision spectroscopy [39].

In the following, we propose an efficient method to generate a GHZ entangled
state of $n$ photons distributed over $n$ microwave cavities that are
coupled by a superconducting \textit{qutrit} (a.k.a. coupler) through
\textit{resonant} interaction. By local operations on a qubit (e.g., an atom
etc.) placed in each cavity, the created GHZ states of photons can be
transferred to qubits for a long time storage and then can be transferred
back to the photons once they are needed to be sent through quantum channels
for implementing quantum communication or quantum information processing in
a network.

As shown below, this proposal does not require measurement on the states of
the coupler qutrit or the cavity-mode photons for each cavity, and only
requires resonant qutrit-cavity interaction and resonant qutrit-pulse
interaction for each step of the operations. Thus, it is relatively
straightforward to implement the method in experiments. Furthermore, the
result of numerical simulation with realistic circuit parameters indicates
that by careful design and optimization high fidelity GHZ states of
multiple cavity photons are within the reach of present day technology.

We emphasize that this proposal is quite general, and can be used to create
GHZ states of photons in multiple cavities with different types of physical
qutrit, such as a Rydberg atom or a quantum dot, as the coupler. Finally, we
show how to apply the method to generate a GHZ state of photons in multiple
cavities using an atom as an example.

The paper is organized as follows. In Sec.~II, we show how to generate a GHZ
state of $n$ photons in $n$ cavities coupled by a superconducting qutrit. In
Sec.~III, we discuss how to extend the method to prepare a GHZ state of $n$
photons in the $n$ cavities using an atom. A concluding summary is given in
Sec.~IV.

\begin{figure}[tbp]
\includegraphics[bb=149 287 555 547, width=8.5 cm, clip]{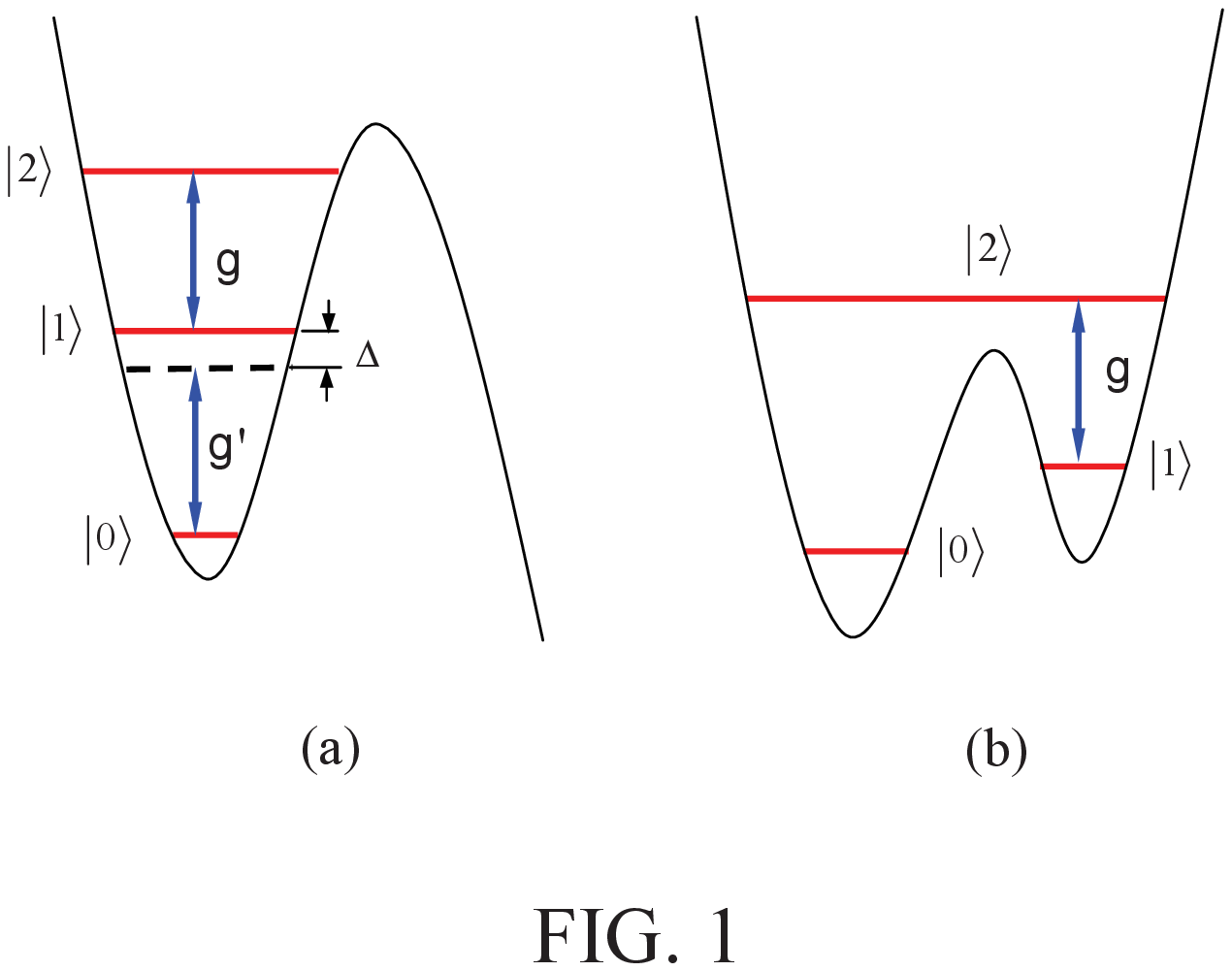} %
\vspace*{-0.08in}
\caption{(Color online) Illustration of qutrit-cavity resonant interaction.
The cavity mode is resonant with the $\left| 1\right\rangle \leftrightarrow
\left| 2\right\rangle $ transition of the qutrit. $g$ is the coupling
constant between the cavity mode and the $\left| 1\right\rangle
\leftrightarrow \left| 2\right\rangle $ transition. In (a), the cavity mode
is decoupled from the $\left| 0\right\rangle \leftrightarrow \left|
1\right\rangle $ transition of a phase qutrit as long as the large detuning
condition $\Delta \gg g^{\prime }$ is satisfied. Here, $\Delta$ is the
detuning between the cavity mode frequency and the $\left| 0\right\rangle
\leftrightarrow \left| 1\right\rangle $ transition frequency, $g^{\prime}$
is the coupling constant between the cavity mode and the $\left|
0\right\rangle \leftrightarrow \left| 1\right\rangle $ transition. In (b),
the dipole matrix element between $\left| 0\right\rangle$ and $\left|
1\right\rangle$ can be made much weaker than that between $\left|
1\right\rangle$ and $\left| 2\right\rangle$ by increasing the barrier height
of the double well potential. Thus the coupling between $\left|
0\right\rangle$ and $\left| 1\right\rangle$ via the cavity mode is
negligible. Note that the coupling strength $g$ may vary when the qutrit
couples with different cavities or resonators. Thus, $g$ is replaced by $g_i$
to denote the coupling strength between the qutrit and cavity $i$ ($%
i=1,2,...,n$).}
\label{fig:1}
\end{figure}

\begin{center}
\textbf{II. GENERATION OF A }$\mathbf{N}$\textbf{-PHOTON GHZ STATE IN THE N
CAVITIES VIA A SUPERCONDUCTING QUTRIT}
\end{center}

In this section, we show how to create a $n$-photon GHZ state in $n$
cavities via a superconducting qutrit, estimate the fidelity of the prepared
GHZ state for $n=2,$ $3$ and $4$, and then end with a brief discussion.

\begin{center}
\textbf{A. Generation of }$n$\textbf{-photon GHZ states in }$n$\textbf{\
cavities}
\end{center}

Consider a superconducting qutrit $A$, which has three levels as depicted in
Fig.~1. The three-level structure in Fig.~1(a) applies to superconducting
phase qutrits [7,33,40] and transmon qutrits [41], while the one in
Fig.~1(b) applies to flux qutrits [42]. In addition, the three-level
structure in Fig.~1(a) or Fig. 1(b) is also available in atoms. The coupler
qutrit $A$ shall have the \textrm{following properties}: (i) for the
three-level structure depicted in Fig. 1(a), transition between the two
lowest levels is highly detuned (decoupled) from the mode of each cavity by
prior adjustment of the level spacings of the qutrit; and (ii) for the
three-level structure depicted in Fig. 1(b), the dipole interaction (i.e.,
matrix element) between the two lowest levels is weak by increasing the
potential barrier between the two levels $\left| 0\right\rangle $ and $%
\left| 1\right\rangle $ [43-45]. Note that for superconducting qutrits, the
level spacings can be rapidly adjusted by varying external control
parameters (e.g., magnetic flux applied to phase, transmon, or flux qutrits,
see e.g. [43-46]).

\begin{figure}[tbp]
\includegraphics[bb=60 383 493 599, width=12.0 cm, clip]{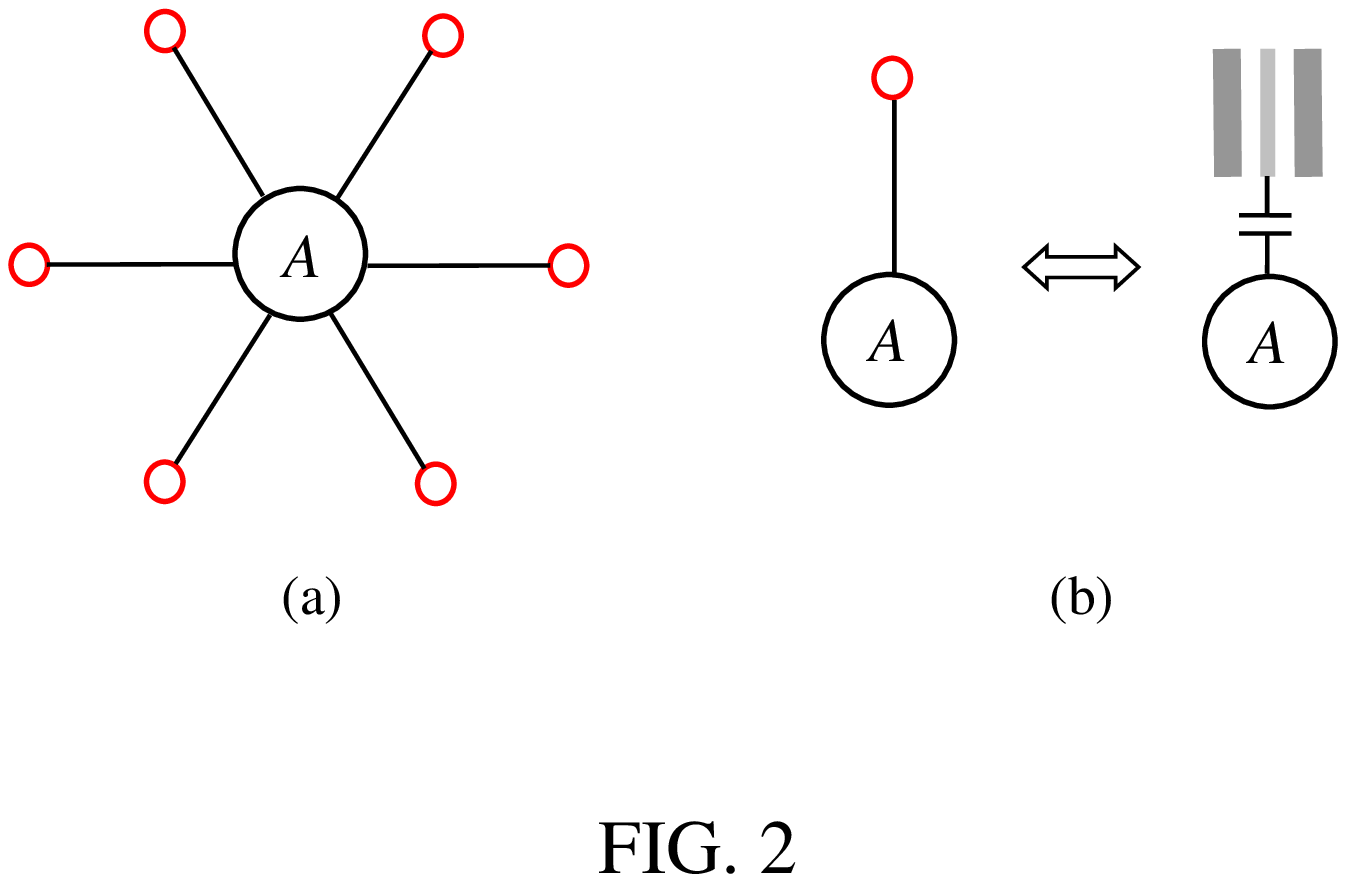} %
\vspace*{-0.08in}
\caption{(Color online) (a) Diagram of a superconducting qutrit $A$ (a
circle at the center) and $n$ cavities. Each red dot represents a
one-dimensional coplanar waveguide resonator which is capacitively coupled
to the coupler qutrit $A$, as shown in (b). (b) The diagram on the left side
is equivalent to the diagram on the right side.}
\label{fig:2}
\end{figure}

Let us now consider $n$ cavities ($1,2,...,n$) each coupled to a
superconducting coupler qutrit $A$ (Fig.~2). Initially, qutrit $A$ is in its
ground state $|0\rangle $ and decoupled from all cavities ($1,2,...,n$) by
prior adjustment of each cavity's frequency; next, qutrit $A$ is transformed
by a $\pi /2$-microwave pulse to the state $\left( \left| 0\right\rangle
+\left| 2\right\rangle \right) /\sqrt{2}$ (hereafter, the three states of
qutrit $A$ are denoted by $\left| 0\right\rangle ,\left| 1\right\rangle ,$
and $\left| 2\right\rangle $ respectively without subscripts) while each
cavity $i$ ($=1,2,...,n$) remains in its vacuum state $\left| 0\right\rangle
_{c,i}$.

To begin with, we define $\omega _{21}$ ($\omega _{20}$) as the $\left\vert
1\right\rangle \leftrightarrow \left\vert 2\right\rangle $ ($\left\vert
0\right\rangle \leftrightarrow \left\vert 2\right\rangle $) transition
frequency of qutrit $A$ and $\Omega _{21}$ ($\Omega _{20}$) as the pulse
Rabi frequency of the coherent $\left\vert 1\right\rangle \leftrightarrow
\left\vert 2\right\rangle $ ($\left\vert 0\right\rangle \leftrightarrow
\left\vert 2\right\rangle $) transition. In addition, the frequency, initial
phase, and duration of the microwave pulse are denoted as \{$\omega ,$ $%
\varphi ,$ $t^{\prime }$\} in the rest of the paper. The operations for
realizing a GHZ state of $n$ photons in the $n$ cavities are described below:

Step $i$ ($i=1,2,...,n-2$): Adjust the frequency $\omega _{c,i}$ of cavity $%
i,$ which will be referred to as the \textit{active} cavity hereafter, such
that it is resonant with the $\left| 1\right\rangle \leftrightarrow \left|
2\right\rangle $ transition of qutrit $A$ (i.e., $\omega _{c,i}=\omega _{21}$%
). After an interaction time $t_i=\pi /(2g_i),$ the state $\left|
0\right\rangle \left| 0\right\rangle _{c,i}$ remains unchanged while the
state $\left| 2\right\rangle \left| 0\right\rangle _{c,i}$ changes to $%
-i\left| 1\right\rangle \left| 1\right\rangle _{c,i}.$ Then, adjust the
frequency of the active cavity away from $\omega _{21}$ to decouple it from
qutrit $A.$ Finally, a microwave pulse of $\{\omega _{21},$ $\pi ,$ $\pi
/\left( 2\Omega _{21}\right) \}$ is \textrm{applied} to qutrit $A$ to
transform its state from $\left| 1\right\rangle $ to $i\left| 2\right\rangle
.$

After executing step $1$ to step $n-2$, the initial state $(\left|
0\right\rangle +\left| 2\right\rangle )\prod_{i=1}^n\left| 0\right\rangle
_{c,i}$ of the whole system is transformed to (here and below a
normalization factor is omitted for simplicity)
\begin{equation}
\left( \left| 0\right\rangle \prod_{i=1}^{n-2}\left| 0\right\rangle
_{c,i}+\left| 2\right\rangle \prod_{i=1}^{n-2}\left| 1\right\rangle
_{c,i}\right) \left| 0\right\rangle _{c,n-1}\left| 0\right\rangle _{c,n}.
\end{equation}

Step $n-1$: Adjust the frequency $\omega _{c,n-1}$ of cavity $n-1$ to have $%
\omega _{c,n-1}=\omega _{21}$ for an interaction time $t_{n-1}=\pi
/(2g_{n-1}).$ As a result, the state $\left\vert 0\right\rangle \left\vert
0\right\rangle _{c,n-1}$ remains unchanged while the state $\left\vert
2\right\rangle \left\vert 0\right\rangle _{c,n-1}$ changes to $-i\left\vert
1\right\rangle \left\vert 1\right\rangle _{c,n-1}$Then, adjust the frequency
of cavity $n-1$ to decouple it from qutrit $A.$ Next, apply a pulse of $%
\{\omega _{20},$ $-\pi /2,$ $\pi /\left( 2\Omega _{20}\right) \}$ to qutrit $%
A$ to transform its state from $\left\vert 0\right\rangle $ to $\left\vert
2\right\rangle $; finally a pulse of $\{\omega _{21},$ $\pi /2,$ $\pi
/\left( 2\Omega _{21}\right) \}$ is applied to qutrit $A$ to transform the
state $\left\vert 1\right\rangle $ to $-\left\vert 2\right\rangle $ and the
state $\left\vert 2\right\rangle $ to $\left\vert 1\right\rangle .$

It is easy to verify that after completing the $n-1$ steps prescribed above,
we obtain the state transformation $\left\vert 0\right\rangle \left\vert
0\right\rangle _{c,n-1}\rightarrow \left\vert 1\right\rangle \left\vert
0\right\rangle _{c,n-1}$ and $\left\vert 2\right\rangle \left\vert
0\right\rangle _{c,n-1}\rightarrow i\left\vert 2\right\rangle \left\vert
1\right\rangle _{c,n-1},$ which propagates state (1) to
\begin{equation}
\left( \left\vert 1\right\rangle \prod_{i=1}^{n-1}\left\vert 0\right\rangle
_{c,i}+i\left\vert 2\right\rangle \prod_{i=1}^{n-1}\left\vert 1\right\rangle
_{c,i}\right) \left\vert 0\right\rangle _{n}.
\end{equation}

Step $n$: Adjust the frequency $\omega _{c,n}$ of cavity $n$ to resonate
with $\omega _{21}$ for an interaction time $t_{n}=\pi /(2g_{n}),$ so that
the state $\left\vert 2\right\rangle \left\vert 0\right\rangle _{c,n}$
changes to $-i\left\vert 1\right\rangle \left\vert 1\right\rangle _{c,n}$
while the state $\left\vert 1\right\rangle \left\vert 0\right\rangle _{c,n}$
remains unchanged. Then, adjust $\omega _{c,n}$ to \textrm{decouple} cavity $%
n$ from qutrit $A.$

It can be seen that after this step of operation, \textrm{state~(2)} becomes

\begin{equation}
\left\vert 1\right\rangle \left( \prod_{i=1}^{n}\left\vert 0\right\rangle
_{c,i}+\prod_{i=1}^{n}\left\vert 1\right\rangle _{c,i}\right) .
\end{equation}
The result~(3) shows that the $n$ cavities are prepared in a $n$-photon GHZ
state $\prod_{i=1}^{n}\left\vert 0\right\rangle
_{c,i}+\prod_{i=1}^{n}\left\vert 1\right\rangle _{c,i},$ while the qutrit $A$
is disentangled from all cavities, after the above $n$-step \textrm{operation%
}.

It should be noticed that rapid tuning of cavity frequencies required by the
proposed protocol has been demonstrated recently in superconducting
microwave cavities (e.g., in less than a few nanoseconds for a
superconducting transmissi\textrm{on line resonator }[47]). Alternatively,
the method can also be implemented with cavities of different resonant
frequencies by rapid tuning of level spacing $\omega _{21}$ of the coupler
qutrit.

Let us now discuss issues which are most relevant to the experimental
implementation of the method. For the method to work the primary
considerations shall be given to:

($a$) The total operation time $\tau ,$ given by
\begin{equation}
\tau =\sum_{i=1}^n\pi /(2g_i)+\left( n-1\right) \pi /\left( 2\Omega
_{21}\right) +\pi /\left( 2\Omega _{20}\right) +2nt_d\text{ }
\end{equation}
(where $t_d$ is the typical time required for adjusting the cavity mode
frequency), needs to be much shorter than the energy relaxation time $T_1$ ($%
T_1^{^{\prime }}$) and dephasing time $T_2$ ($T_2^{^{\prime }}$) of the
level $\left| 2\right\rangle $ ($|1\rangle $) of qutrit $A,$ such that
decoherence caused by energy relaxation and dephasing of qutrit $A$ is
negligible for the operation. Note that $T_1^{^{\prime }}$ and $T_2^{\prime
} $ of qutrit $A$ are comparable to $T_1$ and $T_2,$ respectively. For
instance, $T_1^{\prime }\sim \sqrt{2}T_1$ and $T_2^{^{\prime }}\sim T_2$ for
phase \textrm{qutrits}.

($b$) For cavity $i$ ($i=1,2,...,n$), the lifetime of the cavity mode is
given by $T_{cav}^{i}=\left( Q_{i}/2\pi \nu _{c,i}\right) /\overline{n}_{i},$
where $Q_{i}$ and $\overline{n}_{i}$ are the (loaded) quality factor and the
average photon number of cavity $i$, respectively. For $n$ cavities, the
lifetime of the cavity modes is given by
\begin{equation}
T_{cav}=\frac{1}{n}\min \{T_{cav}^{1},T_{cav}^{2},...,T_{cav}^{n}\},
\end{equation}
which should be much longer than $\tau ,$ such that the effect of cavity
decay is negligible for the operation.

($c$) For step $i$ ($i=1,2,...,n$) of the operation, there exists a qutrit
mediated interaction (crosstalk) between the active cavity and each of the
remaining $n-1$ \textit{idling} cavities (which are not intended to be
involved in the operation). When qutrit $A$ is in the state $\left|
2\right\rangle ,$ the probability of exciting an idling cavity $j\neq i$
from the vacuum state $|0\rangle _{c,j}$ to $|1\rangle _{c,j},$ after the
completion of step $i$, is \textrm{given} approximately by
\begin{equation}
p_j\approx \frac 12\left( 1-\cos \frac{\pi \sqrt{4\widetilde{g}_j^2+\Delta
_j^2}}{2g_i}\right) \left( 1-\frac{\Delta _j^2}{4\widetilde{g}_j^2+\Delta
_j^2}\right) ,  \label{pj}
\end{equation}
where $\widetilde{g}_j$ is the off-resonant coupling constant between cavity
$j$ and the $\left| 1\right\rangle \leftrightarrow \left| 2\right\rangle $
transition of qutrit $A,$ and $\Delta _j=\omega _{21}-\widetilde{\omega }%
_{c,j}$ is the detuning of the frequency of cavity $j$ with the $\left|
1\right\rangle \leftrightarrow \left| 2\right\rangle $ transition frequency.
Hereafter, $\widetilde{\omega }_{c,j}$ represents the
frequency of cavity $j$ when \textrm{idling} [see Fig. 3(a)].

It can be seen from Eq. (\ref{pj}) that $p_{j}$ is negligibly small when $%
\Delta _{j}\gg \widetilde{g}_{j}$. Hence, as long as the large detuning
condition is \textrm{satisfied} for all of the idling cavities, crosstalk
caused error can be suppressed to a \textrm{tolerable} level.

($d$) For step $i$ ($i=1,2,...,n$) of the operation, there also
exists an inter-cavity cross coupling which is determined
mostly by the coupling capacitance $C_c$ and the qutrit's self capacitance $%
C_q$, because field leakage through space is extremely low for
high-$Q$ cavities as long as inter-cavity distances are much greater
than transverse dimension of the cavities - a condition easily met in
experiments for $n\leq 8$. Furthermore, as the result of our numerical
simulation shown below (see Fig. 4), {\normalsize the effects of these
inter-cavity couplings can however be made negligible as long as }$%
g_{kl}\leq 10^{-2}g_i,${\normalsize \ where }$g_{kl}$ is \textrm{the
corresponding inter-cavity coupling constant between cavities }$k$ and $l$.
\begin{figure}[tbp]
\includegraphics[bb=80 208 568 691, width=8.5 cm, clip]{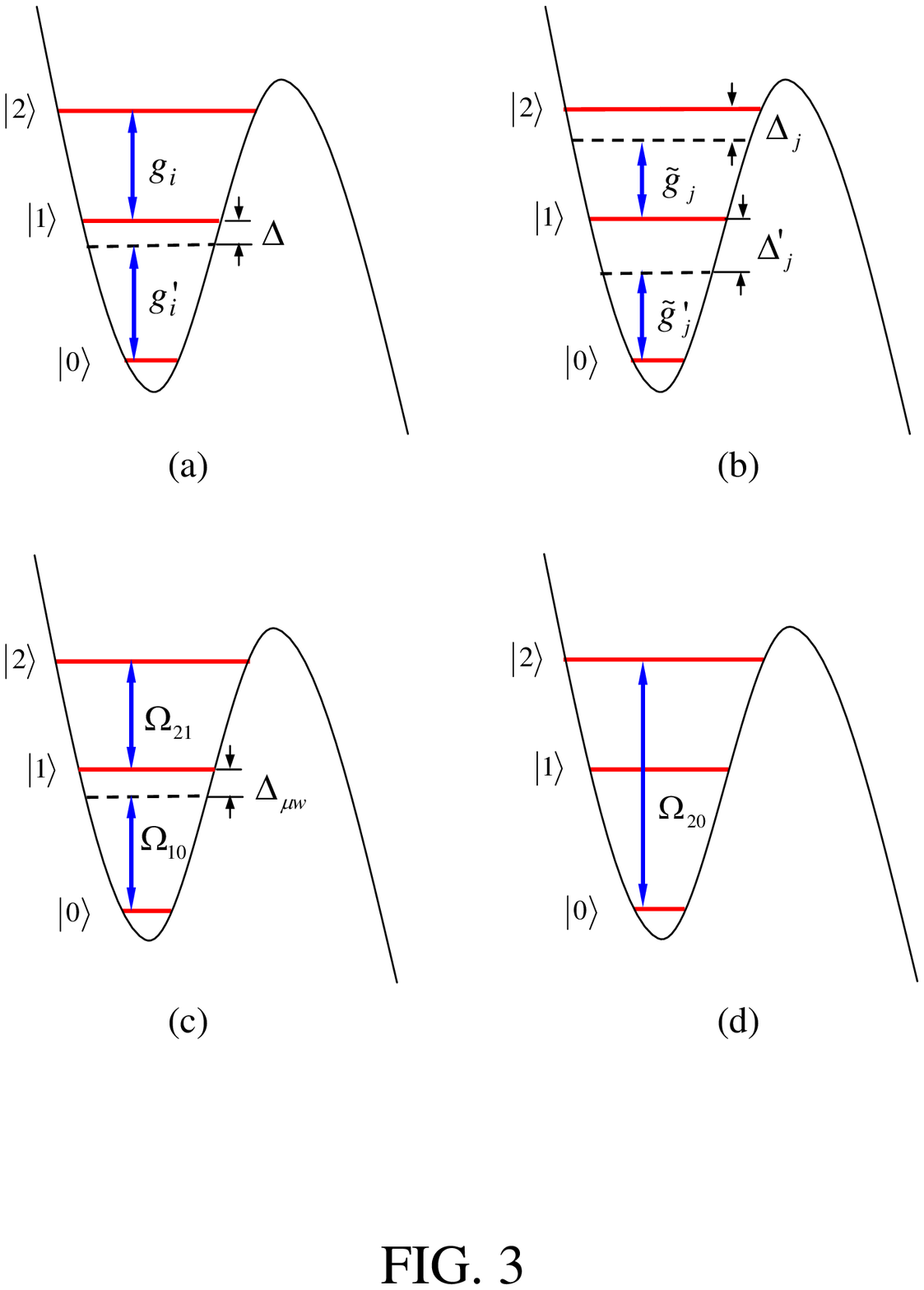} %
\vspace*{-0.08in}
\caption{(Color online) Illustration of qutrit-cavity or qutrit-pulse
interaction. (a) Cavity $i$ is resonant with the $%
\left| 1\right\rangle \leftrightarrow \left| 2\right\rangle $ transition of
qutrit $A$ when $\omega _{c,i}=\omega _{21}$ with a coupling constant $g_i$
but off-resonant with the $\left| 0\right\rangle \leftrightarrow \left|
1\right\rangle $ transition with a coupling constant $g_i^{\prime }$ and
detuning $\Delta =\omega _{10}-\omega _{c,i}.$ (b) Cavity $j$ of frequency
$\widetilde{\omega }_{c,j}$ is off-resonant with
the $\left| 1\right\rangle \leftrightarrow \left|
2\right\rangle $ ($\left| 0\right\rangle \leftrightarrow \left|
1\right\rangle $) transition of qutrit $A$ with a coupling constant $%
\widetilde{g}_j$ ($\widetilde{g}_j^{\prime }$) and detuning $\Delta
_j=\omega _{21}-\widetilde{\omega }_{c,j}$ ($\Delta _j^{\prime }=\omega
_{10}-\widetilde{\omega }_{c,j}$)$.$  (c) Represents the situation
when a microwave classical pulse of frequency $\omega =\omega _{21}$ is
applied to qutrit $A$ but off-resonant with the $\left| 0\right\rangle
\leftrightarrow \left| 1\right\rangle $ transition with detuning $\Delta
_{\mu w}=\omega _{10}-\omega .$ The corresponding Rabi frequencies are $%
\Omega _{21}$ and $\Omega _{10},$ respectively. (d) A microwave pulse of
frequency $\omega =\omega _{20}$ is applied to qutrit $A$ with the
corresponding Rabi frequency $\Omega _{20}.$ Note that for (c), the coupling
of the pulse to the $\left| 0\right\rangle \leftrightarrow \left|
2\right\rangle $ transition is negligible due to the fact that the pulse is
highly detuned from the $\left| 0\right\rangle \leftrightarrow \left|
2\right\rangle $ transition frequency. For the same reason, for (d), the
coupling of the pulse to the $\left| 0\right\rangle \leftrightarrow \left|
1\right\rangle $ transition and the $\left| 1\right\rangle \leftrightarrow
\left| 2\right\rangle $ transitions is negligible as well.}
\label{fig.}
\end{figure}

\begin{center}
\textbf{B. Fidelity}
\end{center}

The proposed protocol for creating the $n$-photon GHZ state described above
involves three basic types of transformation:

(i) The first one requires that during step $i$ ($i=1,2,...,n$) of the
operation, cavity $i$ is tuned to resonant with the $\left| 1\right\rangle
\leftrightarrow \left| 2\right\rangle $ transition of qutrit $A$ while other
cavities are decoupled from qutrit $A.$ In the interaction picture (the same
without mentioning hereafter), the interaction Hamiltonian governing this
basic transformation is given by
\begin{eqnarray}
H_{I,1} &=&g_i\left( a_iS_{12}^{+}+h.c.\right) +g_i^{\prime }\left(
e^{i\Delta t}a_iS_{01}^{+}+h.c.\right)  \nonumber \\
&&\ +\sum_{j\neq i,j=1}^n\widetilde{g}_j\left( e^{i\Delta
_jt}a_jS_{12}^{+}+h.c.\right) +\sum_{j\neq i,j=1}^n\widetilde{g}_j^{\prime
}\left( e^{i\Delta _j^{\prime }t}a_jS_{01}^{+}+h.c.\right)  \nonumber \\
&&+\sum_{k\neq l;k,l=1}^ng_{kl}\left( e^{i\Delta
_{kl}t}a_ka_l^{+}+h.c.\right) .
\end{eqnarray}
where $S_{12}^{+}=\left| 2\right\rangle \left\langle 1\right| ,$ $%
S_{01}^{+}=\left| 1\right\rangle \left\langle 0\right| ,$ and $a^{+}$($a$)
is the cavity photon creation (annihilation) operator. The first term
describes the resonant coupling between cavity $i$ and the $\left|
1\right\rangle \leftrightarrow \left| 2\right\rangle $ transition of qutrit $%
A$ with a coupling constant $g_i$ [Fig. 3(a)] while the second term
represents the off-resonant coupling between cavity $i$ and the $\left|
0\right\rangle \leftrightarrow \left| 1\right\rangle $ transition with a
coupling constant $g_i^{\prime }$ and detuning $\Delta =\omega _{10}-\omega
_{c,i}$ [Fig. 3(a)]. The third (fourth) term is the off-resonant coupling
between all idling cavities and the $\left| 1\right\rangle \leftrightarrow
\left| 2\right\rangle $ ($\left| 0\right\rangle \leftrightarrow \left|
1\right\rangle $) transition$,$ where $\widetilde{g}_j$ ($\widetilde{g}%
_j^{\prime }$) is the coupling constant between cavity $j$ and the $\left|
1\right\rangle \leftrightarrow \left| 2\right\rangle $ ($\left|
0\right\rangle \leftrightarrow \left| 1\right\rangle $) transition, with
detuning $\Delta _j=\omega _{21}-\widetilde{\omega }_{c,j}$ ($\Delta
_j^{\prime }=\omega _{10}-\widetilde{\omega }_{c,j}$) [Fig. 3(b)]. The last
term represents the inter-cavity crosstalk between any two cavities $k$ and $%
l$, where $\Delta _{kl}$ is the frequency detuning for the two cavities $k$
and $l.$

(ii) The second one involves pulse-qutrit interaction by applying a
microwave pulse (with frequency $\omega =\omega _{21}$ and initial phase $%
\varphi $) to qutrit $A$. Note that when the pulse is on, all cavities are
required to be decoupled from qutrit $A$ by a prior detuning of their
frequencies from $\omega _{21}.$ The interaction Hamiltonian for this basic
transformation is \textrm{given by}
\begin{eqnarray}
H_{I,2} &=&\Omega _{21}\left( e^{-i\varphi }S_{12}^{+}+h.c.\right) +\Omega
_{10}\left[ e^{i(\Delta _{\mu w}t-\varphi )}S_{01}^{+}+h.c.\right]  \nonumber
\\
&&+\sum_{j=1}^n\widetilde{g}_j\left( e^{i\Delta
_jt}a_jS_{12}^{+}+h.c.\right) +\sum_{j=1}^n\widetilde{g}_j^{\prime }\left(
e^{i\Delta _j^{\prime }t}a_jS_{01}^{+}+h.c.\right)  \nonumber \\
&&+\sum_{k\neq l;k,l=1}^ng_{kl}\left( e^{i\Delta
_{kl}t}a_ka_l^{+}+h.c.\right) ,
\end{eqnarray}
where $\Omega _{10}$ is the pulse Rabi frequency associated with the $\left|
0\right\rangle \leftrightarrow \left| 1\right\rangle $ transition, and $%
\Delta _{\mu w}=\omega _{10}-\omega $ is the detuning between the pulse
frequency $\omega $ and the $\left| 0\right\rangle \leftrightarrow \left|
1\right\rangle $ transition frequency $\omega _{10}$ [Fig. 3(c)].

(iii) The last one requires that during the operation of step $n$ (the final
step operation above), a microwave pulse (with frequency $\omega =\omega
_{20}$ and initial phase $\varphi $) is applied to qutrit $A$ while each
cavity is decoupled from qutrit $A$. The interaction Hamiltonian governing
this basic transformation is given by
\begin{equation}
H_{I,3}=\Omega _{20}\left( e^{-i\varphi }S_{02}^{+}+h.c.\right) +\varepsilon
,
\end{equation}
where $\varepsilon $ is the sum of the last three terms of Eq. (8), $%
S_{02}^{+}=\left| 2\right\rangle \left\langle 0\right| $, and the terms
describing the pulse induced coherent $\left| 0\right\rangle \leftrightarrow
\left| 1\right\rangle $ and $\left| 1\right\rangle \leftrightarrow \left|
2\right\rangle $ transitions are negligible because $\omega \gg \omega
_{10},\omega _{21}$ [Fig. 3(d)].

For each of the three basic types of transformation described above, the dynamics
of the lossy system, composed of all cavities and qutrit $A$, is determined
by

\begin{eqnarray}
\frac{d\rho }{dt} &=&-i\left[ H_{I},\rho \right] +\sum_{i=1}^{n}\kappa _{i}%
\mathcal{L}\left[ a_{i}\right] +\left\{ \gamma _{\varphi ,21}\left(
S_{21}^{z}\rho S_{21}^{z}-\rho \right) +\gamma _{21}\mathcal{L}\left[
S_{21}^{-}\right] \right\}  \nonumber \\
&&\ \ +\left\{ \gamma _{\varphi ,20}\left( S_{20}^{z}\rho S_{20}^{z}-\rho
\right) +\gamma _{20}\mathcal{L}\left[ S_{20}^{-}\right] \right\} +\left\{
\gamma _{\varphi ,10}\left( S_{10}^{z}\rho S_{10}^{z}-\rho \right) +\gamma
_{10}\mathcal{L}\left[ S_{10}^{-}\right] \right\} ,
\end{eqnarray}
where $H_{I}$ is the $H_{I,1},$ $H_{I,2}$ or $H_{I,3}$ above, $\mathcal{L}%
\left[ a_{i}\right] =a_{i}\rho a_{i}^{+}-a_{i}^{+}a_{i}\rho /2-\rho
a_{i}^{+}a_{i}/2,$ $\mathcal{L}\left[ S_{ij}^{-}\right] =S_{ij}^{-}\rho
S_{ij}^{+}-S_{ij}^{+}S_{ij}^{-}\rho /2-\rho S_{ij}^{+}S_{ij}^{-}/2$ ($%
ij=21,20,10$), $S_{21}^{z}=\left\vert 2\right\rangle \left\langle
2\right\vert -\left\vert 1\right\rangle \left\langle 1\right\vert $, $%
S_{20}^{z}=\left\vert 2\right\rangle \left\langle 2\right\vert -\left\vert
0\right\rangle \left\langle 0\right\vert $, and $S_{10}^{z}=\left\vert
1\right\rangle \left\langle 1\right\vert -\left\vert 0\right\rangle
\left\langle 0\right\vert $. In addition, $\kappa _{i}$ is the decay rate of
the mode of cavity $i,$ $\gamma _{\varphi ,21}$ ($\gamma _{\varphi ,20}$)
and $\gamma _{21}$ ($\gamma _{20}$) are the dephasing rate and the energy
relaxation rate of the level $\left\vert 2\right\rangle $ of qutrit $A$ for
the decay path $\left\vert 2\right\rangle \rightarrow \left\vert
1\right\rangle $ ($\left\vert 0\right\rangle $), respectively and $\gamma
_{\varphi ,10}$ and $\gamma _{10}$ are those of the level $\left\vert
1\right\rangle $ for the decay path $\left\vert 1\right\rangle \rightarrow
\left\vert 0\right\rangle $. The fidelity of the operation is given by
\begin{equation}
\mathcal{F}=\left\langle \psi _{id}\right\vert \widetilde{\rho }\left\vert
\psi _{id}\right\rangle ,
\end{equation}
where $\left\vert \psi _{id}\right\rangle $ is the state (3) of an ideal
system (i.e., without dissipation, dephasing, and crosstalks) and $%
\widetilde{\rho }$ is the final density operator of the system when the
operation is performed in a realistic physical \textrm{system}.

\begin{figure}[tbp]
\begin{center}
\includegraphics[bb=0 0 784 488, width=16.5 cm, clip]{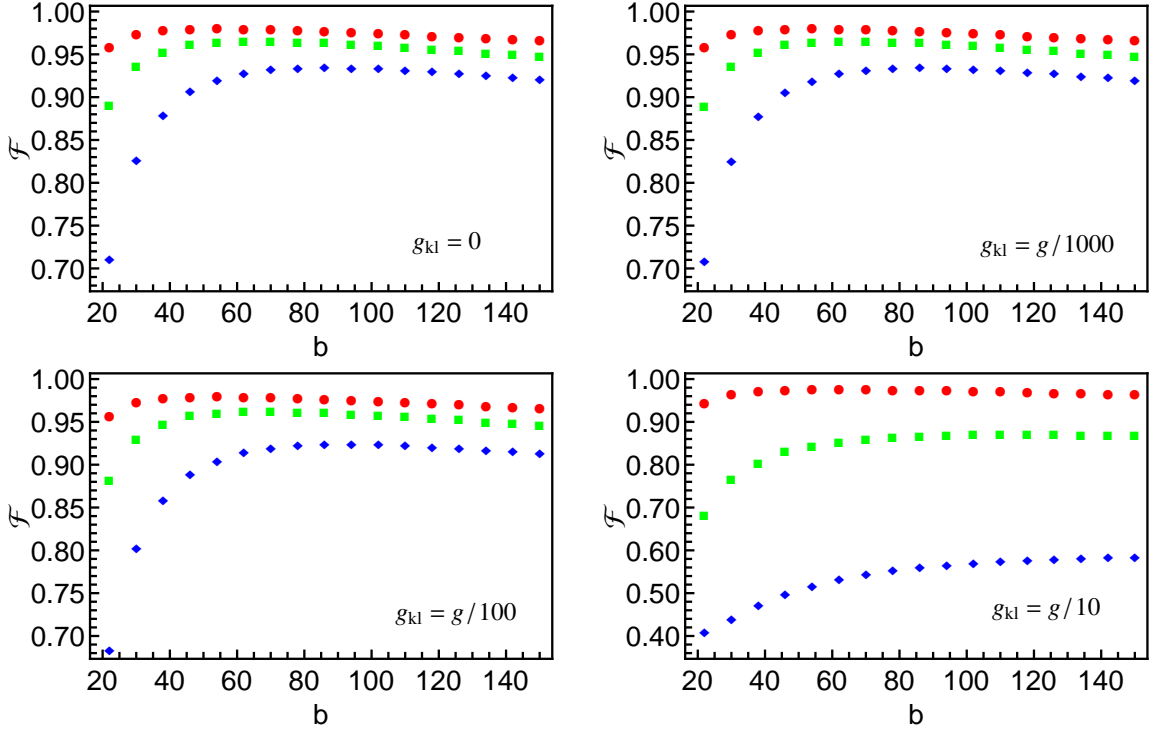} %
\vspace*{-0.08in}
\end{center}
\caption{(Color online) Fidelity versus $b=\Delta/g^{\prime}$. Refer to the
text for the parameters used in the numerical calculation. Here, $g_{kl}$ is
the coupling strength between cavities $k$ and $l$ ($k\neq l;$ and $%
k,l=1,2,3,4$), which are taken to be the same for simplicity. In each
figure, the red, green, and blue lines correspond to $n=2, 3,$ and $4$,
respectively.}
\label{fig:4}
\end{figure}

We now numerically calculate the fidelity of the prepared GHZ state of
photons in up to four cavities. Without loss of generality, let us consider
a phase qutrit with {\normalsize three} levels in the metastable potential
well, for which $\omega _{10}/2\pi \sim 6.8$ GHz and $\omega _{21}/2\pi \sim
6.3$ GHz [33]. The frequency $\omega _{c,i}/2\pi $ of the \textit{active}
cavity $i$ ($i=1,2,3,4$) is thus $\sim 6.3$ GHz, resulting in $\Delta /2\pi $
$\sim 500$ MHz. For the \textit{idling} cavity $j$ ($j=1,2,3,4$),
{\normalsize we} choose $\widetilde{\omega }_{c,j}/2\pi $ $\sim 5.6$ GHz
[47], which leads to $\Delta _j/2\pi $ $\sim 700$ MHz and $\Delta _j^{\prime
}/2\pi \sim 1.2$ GHz. For the phase qutrit here, one has $g_i\sim \sqrt{2}%
g_i^{\prime },$ $\widetilde{g}_j\sim \sqrt{2}\widetilde{g}_j^{\prime }$ and $%
\widetilde{g}_j\sim g_i\sqrt{\widetilde{\omega }_{c,j}/\omega _{c,j}}%
(i,j=1,2,3,4$). For simplicity, assume that $g_1=g_2=g_3=g_4\equiv g$ and
thus $g_1^{\prime }=g_2^{\prime }=g_3^{\prime }=g_4^{\prime }\equiv
g^{\prime }.$ Other parameters used in the numerical calculation are as
follows: (i) $\Delta _{\mu w}/2\pi =500$ MHz, $\Omega _{21}\sim \sqrt{2}%
\Omega _{10},$ $\Omega _{10}/2\pi =50$ MHz, and $\Omega _{20}/2\pi $ $=200$
MHz (\textrm{which is available in experiments [48]}), (ii) $\gamma
_{\varphi ,21}^{-1}=\gamma _{\varphi ,20}^{-1}=\gamma _{\varphi ,10}^{-1}=5$
$\mu $s, $\gamma _{21}^{-1}=25$ $\mu $s, $\gamma _{20}^{-1}=200$ $\mu $s
[49], $\gamma _{10}^{-1}=50$ $\mu $s, $\kappa _1^{-1}=\kappa _2^{-1}=\kappa
_3^{-1}=\kappa _4^{-1}=20$ $\mu $s. For the parameters chosen here, the
fidelity versus $b\equiv \Delta /g^{\prime }$ is shown in Fig.~4, from which
one can see that for $b=50,$ $60$ and $85,$ a high fidelity $\sim 98\%,$ $%
97\%,$ and $93\%$ can be respectively achieved for $n=2,$ $3,$ and $4$ when $%
g_{kl}\leq g/100$ ($k\neq l;$ and $k,l=1,2,3,4$). Interestingly, it is noted
from Fig. 4 that the effect of direct coupling between cavities on the
fidelity of the prepared GHZ states is negligible when the {\normalsize inter%
}-cavity coupling strength ($g_{kl}$) is smaller than $g$ by two orders of
magnitude. This condition, $g_{kl}/g\leq 0.01$, is not difficult to satisfy
with typical capacitive cavity-qutrit coupling illustrated in Fig. 2(b). In
this case, because very little field could leak out of each cavity it can
be shown that as long as the cavities are physically well separated, the
inter-cavity crosstalk coupling strength {\normalsize is} $g_{kl}\approx
g(C_c/C_\Sigma ),$ \textrm{where }$C_c\sim 1$\textrm{\ fF and }$C_\Sigma
=nC_c+C_q\sim 10^2$\textrm{\ fF are the typical value of the cavity-qutrit
coupling capacitance and the sum of all coupling capacitance and qutrit self
capacitance, respectively. }Therefore, it is straightforward to implement
designs with {\normalsize sufficiently} weak direct inter-cavity couplings.

Let us focus on the case of four cavities. For $b=85$, we have $g/2\pi \sim
8.3$ MHz, $g^{\prime }/2\pi \sim 5.9$ MHz, $\widetilde{g}_j/2\pi $ $\sim 7.8$
MHz, and $\widetilde{g}_j^{\prime }\sim 5.5$ GHz ($j=1,2,3,4$). Note that a
qutrit-cavity coupling constant $g/2\pi \sim 220$ MHz can be reached for a
superconducting qutrit coupled to a one-dimensional standing-wave CPW
(coplanar waveguide) resonator [30], and that $T_1^{\prime }$ and $%
T_2^{\prime }$ can be made to be a few tens of $\mu $s for the state of art
superconducting qutrits at the present time [50]. For the cavity resonant
frequency $\sim 6.3$ GHz chosen here and for the $\kappa _1^{-1},\kappa
_2^{-1},\kappa _3^{-1},$ $\kappa _4^{-1}$ used in the numerical calculation,
the required quality factor for the four cavities is $Q\sim 7.9\times 10^5$.
Note that superconducting CPW resonators with a loaded quality factor $Q\sim
10^6$ have been experimentally demonstrated [51,52], and planar
superconducting resonators with internal quality factors above one million ($%
Q>10^6$) have also been reported recently [53]. Our analysis given here
demonstrates that preparation of the GHZ state of photons in up to four
cavities is feasible within the present circuit QED \textrm{technique}.

Before ending this subsection, we point out that the non-monotonic
dependence of fidelity $\mathcal{F}$ on the dimensionless parameter $b$
observed in Fig. 4 are essentially an artifact of the numerical procedure.
In our numerical calculation, $b$ is on the increase by keeping the detuning
$\Delta $ ($\sim 500$ MHz) constant while reducing $g^{\prime }$ which
corresponds to decreasing the qutrit-cavity coupling capacitance $C_c$.
Since the ratio $g/g^{\prime }$ is determined by the qutrit's level
structure and thus remains constant irrespective the value of coupling
capacitance $C_c,$ the protocol would{\normalsize \ thus} take a longer time
to complete as $g^{\prime }$, and thus $g,$ is reduced to a value below
which the {\normalsize adverse effects of} cavity decay and \textrm{qutrit
decoherence }{\normalsize take over}$.$

\begin{center}
\textbf{C. Discussion}
\end{center}

In principle, the method presented above can be used to create a GHZ state
of $n$ photons in $n$ cavities. However, it should be pointed out that in
the solid-state setup {\normalsize scaling up} to many cavities coupled to a
single superconducting qutrit will introduce new challenges. For instance,
the coupling constant between the coupler qutrit $A$ and each cavity
decreases as the number of cavities increases. As a result, the operation
becomes slower and thus decoherence, caused due to qutrit-environment
interaction and/or cavity decay, may become a severe problem. Since $g_i$ is
inversely proportional to $n,$\ the number of cavities coupled to qutrit $A$%
\ may be limited to about 4 to 6 to maintain sufficiently strong
qutrit-cavity couplings.

Tunable resonators usually come with a non-linearity [54,55]. Details on how
to tune the frequency of a resonator can be found in Refs. [54,55]. We
remark that how to tune frequency of a resonator is not the main focus of
this paper, which is beyond the scope of this theoretical work. In addition,
the energy relaxation time of qutrit $A$ can be shortened by the Purcell
decay of the resonators, which however can be made negligible with a high-$Q$
resonator [56]. A detailed discussion on this issue is out of the scope of
this work.

It should be mentioned that three-level superconducting qutrits were earlier
used for quantum operations within cavity QED [10,18,19]. We stress that the
present work is quite different from the previous one [33]. As discussed in
[33], the NOON state of the two resonators was created by first preparing a
Bell state of two superconducting qutrits (connecting to the two resonators
separately) and then swapping the prepared Bell state of the two qutrits to
the two resonators. Thus, if the protocol in [33] is applied to generate a
GHZ state of $n$ cavities, one will need to first prepare a GHZ state of $n$
superconducting qubits (each connecting to a resonator) and then swap the
prepared GHZ state of the $n$ qubits to the $n$ cavities. However, as shown
above, prior preparation of a GHZ state of $n$ superconducting qubits is not
required by the present proposal. Moreover, by using the protocol in [33] to
implement the current task, $n$ superconducting qubits are required; while
only a coupler qutrit $A$ is needed by the present proposal.

\begin{center}
\textbf{III. GENERATION OF A }$\mathbf{N}$\textbf{-PHOTON GHZ STATE IN THE N
CAVITIES USING AN ATOM}
\end{center}

During the past decade, much attention has been paid to the generation of
highly entangled states with atomic systems. Two-atom entangled states and
three-particle GHZ entangled states (with two atoms plus one cavity mode)
have been experimentally demonstrated in microwave cavity QED [4,5]. In
addition, based on cavity QED, numerous theoretical proposals have been
presented for entangling atoms coupling to the mode (s) of a single cavity
[57] and atoms in two or more cavities [58]. In principle, an entangled
state of $n$ photons in $n$ cavities ($n\geq 2$) can be created, by first
preparing an $n$-atom entangled state using the previous proposals [57,58],
and then transferring the prepared $n$-atom entangled states onto $n$
photons in the $n$ cavities via the state transfer from an atom to a photon
in a cavity. In the following, we will present an alternative way to
implement an $n$-photon GHZ state, which, as shown below, does not require
prior preparation of atomic entangled states. The scheme presented here is
actually a generalization of the method described in Sec. II to GHZ-state
generation of photons in multiple cavities through an atom.

\begin{figure}[tbp]
\includegraphics[bb=32 445 580 581, width=15.0 cm, clip]{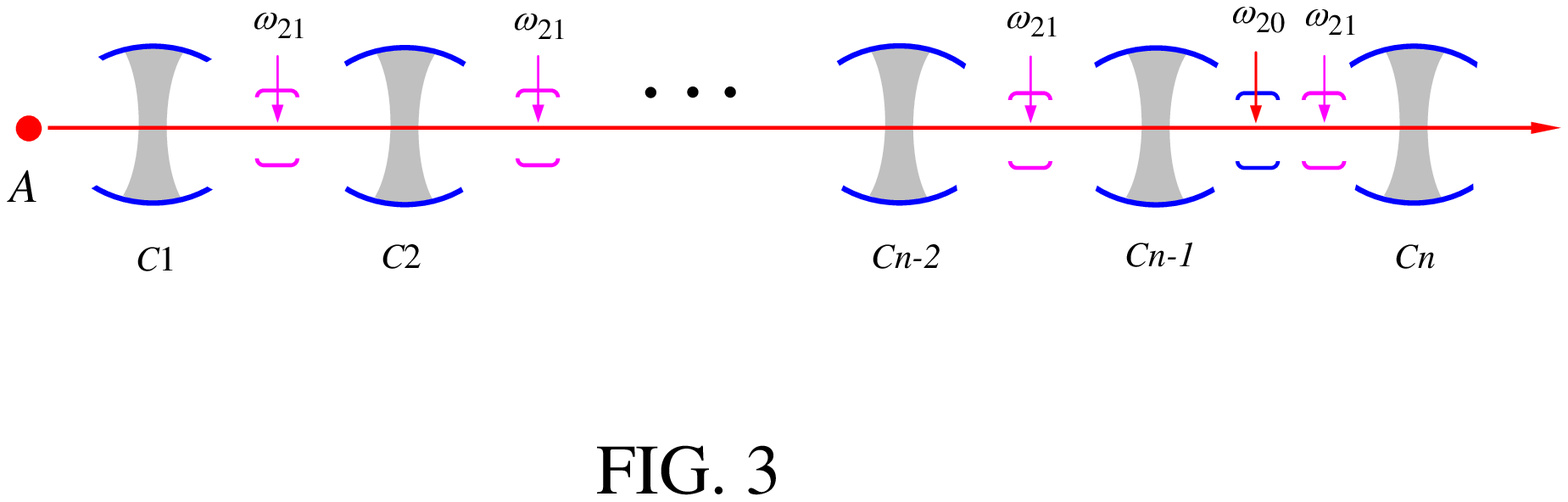} %
\vspace*{-0.08in}
\caption{(Color online) Diagram of $n$ identical cavities and an atom $A$ (a
red dot). The atom $A$ is sent through or moved into each cavity for an
interaction time $\pi /\left( 2g\right) .$ Before arriving in cavity $n-1,$
the atom $A$ is addressed by a classical pulse (with frequency $\omega =$ $%
\omega _{21},$ initial phase $\pi ,$ and duration $\pi /\left( 2\Omega
_{21}\right) $) after it leaves each cavity (see the pink-color frame with
an arrow). When the atom $A$ exits the cavity $n-1$, two pulses are applied
to it. The first pulse has frequency $\omega =$ $\omega _{20},$ initial
phase $-\pi /2,$ and duration $\pi /\left( 2\Omega _{20}\right) $ (see the
blue-color frame with an arrow) while the second pulse has frequency $\omega
=\omega _{21},$ initial phase $\pi /2,$ and duration $\pi /\left( 2\Omega
_{21}\right) $. Here, $g$ is the coupling constant between the cavity mode
and the $\left| 1\right\rangle \leftrightarrow \left| 2\right\rangle $
transition of the atom $A;$ $\omega _{20}$ and $\omega _{21}$ are the $%
\left| 2\right\rangle \leftrightarrow \left| 0\right\rangle $ transition
frequency and the $\left| 2\right\rangle \leftrightarrow \left|
1\right\rangle $ transition frequency of the atom $A,$ respectively. In
addition, $\Omega_{21}$ ($\Omega_{20}$) is the Rabi frequency of the pulse
associated with the $\left| 1\right\rangle \leftrightarrow \left|
2\right\rangle $ transition ($\left| 0\right\rangle \leftrightarrow \left|
2\right\rangle $ transition) of the atom $A$.}
\label{fig:5}
\end{figure}

Consider $n$ identical cavities ($1,2,...,n$) and an atom $A$ with three
levels as depicted in Fig. 1. The atom $A$ is initially prepared in the
state $\left( \left| 0\right\rangle +\left| 2\right\rangle \right) /\sqrt{2}$
and each cavity is in a vacuum state, i.e., $\left| 0\right\rangle _{c,i}$
for cavity $i$ ($i=1,2,...,n$). In addition, assume that the cavity mode of
each cavity is resonant with the $\left| 1\right\rangle \leftrightarrow
\left| 2\right\rangle $ transition but highly detuned (decoupled) from the
transition between any other two levels of the atom $A.$ The procedure for
generating a GHZ state of $n$ photons in the $n$ cavities is illustrated in
Fig.~3. The total operation time $\tau $ is given in Eq.~(4), in which $\tau
_d$ is now a typical time for moving atom $A$ into or out of a cavity. The
number of cavities to be prepared in an entangled state is limited by the
decay of atom $A$ and decay of each cavity.

The present scheme has the following advantages: (i) Only one atom is
needed; (i) Neither measurement on the states of the atom $A$ nor
measurement on the cavity photons is needed; (ii) No adjustment of the
atomic level spacings or the cavity mode frequency is needed during the
entire operation.

We should mention that the atom-cavity interaction time can be tuned by
changing the atomic velocity in the case when the atom $A$ is sent through
each cavity [59]. In addition, it can be tuned by controlling the duration
of the atom in each cavity, for the case when the atom is loaded into or out
of a cavity by trapping the atom in a linear trap [60], inside an optical
lattice [61], or on top of an atomic chip [62]. Note that the approach for
trapping and moving atoms into or out of a cavity has been employed in the
earlier work for quantum computing with atoms in cavity QED [63-66].

To investigate the experimental feasibility of this scheme, let us consider
preparation of a GHZ state for 10 photons in ten cavities using a single
Rydberg atom. The atom $A$ is chosen as a Rydberg atom with principal
quantum numbers 50 and 51 (respectively corresponding to the levels $\left|
1\right\rangle $ and $\left| 2\right\rangle $). For the Rydberg atom chosen
here, the $\left| 1\right\rangle \leftrightarrow \left| 2\right\rangle $
transition frequency is $\omega _{21}/2\pi \sim 51.1$ GHz [67], the coupling
constant is $g=2\pi \times 50$ KHz [68], the energy relaxation time of the
level $\left| 2\right\rangle $ is $T_r\sim 3\times 10^{-2}$ s [69], and the
dephasing time $T_\varphi \sim 10^{-3}$ s of the level $\left|
2\right\rangle $ can be reached in the present experiment [70]. With the
choice of $t_d\sim 1\mu $s and $\Omega _{21}\sim \Omega _{20}\sim 10g,$ we
have $\tau \sim 7.5\times 10^{-5}$ s $\ll T_r,T_\varphi .$

In the present case, the mode frequency of each cavity is $\sim 51.1$ GHz.
One can see from the above discussion that each cavity was occupied by a
single photon during the GHZ-state preparation. For a cavity with $%
Q=10^{10}, $ we have $\min \{T_{cav}^1,T_{cav}^2,...,T_{cav}^{10}\}\sim
3.1\times 10^{-2}$ s, resulting in $T_{cav}\sim 3.1\times 10^{-3}$ s for $%
n=10,$ which is much longer than $\tau .$ Note that cavities with a high $%
Q\sim 3\times 10^{10}$ was previously reported [71]. Thus, generating a GHZ
state of 10 photons in ten cavities with assistance of an atom is possible
within the present cavity QED technique.

By Using \textit{linear optics elements} and \textit{single photon detectors}%
, many schemes for creating entangled multi-photon states have also been
proposed [72]; and experimental realization of an eight-photon GHZ state [1]
and a three-photon W state [73] has been reported. However, this type of
approaches is much more difficult to implement than cavity QED for hybrid
systems consisting of photons and matter qubits of nature made and/or
engineered. The present work represents a significant advancement in circuit
and atom QED because it provides a simple and fast approach for
deterministically creating a multi-photon GHZ state, which needs only a
single coupler qubit and does not require measurement or detection on
photons.

We noticed that two previous works [74,75] are relevant to ours. Ref.~[74]
presents a scheme for preparation of a GHZ-type entangled \textit{coherent
state} of $n$ cavities by having an atom interacts with each of the cavities
dispersively and then measuring the state of the atom. We are aware of that
a GHZ entangled Fock state of photons in multiple cavities can in principle
be generated using the same procedure described in [74]. However, the method
has the following drawbacks: (i) the operation is rather slow because of the
dispersive atom-cavity interaction, (ii) a measurement on the state of the
atom is required, and (iii) since the prepared GHZ state depends on the
measurement outcome on the atomic states, the GHZ-state preparation is not
deterministic. In contrast, our proposal mitigates these problems
effectively: the operation is much faster because of the resonant
atom-cavity interactions; there is no need to measure the state of the atom;
and the generation of the GHZ state is deterministic. Ref.~[75] proposes a
method for preparing a cluster state of photons in $n$ cavities via resonant
atom-cavity interactions. However, our proposal is significantly different
from that of [75]. First, we focus on preparing a GHZ entangled Fock state
of photons in multiple cavities. Second, an $n$-qubit cluster state cannot
be transformed into a GHZ state (for $n>3$)~[76]. Last, the method proposed
in [75] requires an atom to interact with \textit{two} classical pulses
after it leaves each cavity (except the final one) while our proposal only
requires the atom interacting with \textit{one} classical pulse after it
exits each cavity (except the final one).

After a thorough search, we found that three schemes [77-79] were previously
proposed for implementing the GHZ state of photons in $n$ cavities by
sending an atom through $n$ cavities. However, these schemes require
measuring the state of the atom and/or using $n$ levels of the atom (i.e.,
the number of the atomic levels used needs to be equal to the number of the
cavities).

Finally, our work is different from the previous one in [80], in which a
matrix-product state (i.e., a generalized version of the GHZ state) was
produced through sequential interaction between atomic and photonic qubits.
In [80], the authors discussed how to create different entangled states of
photons at the output of a cavity, while in our case we consider how to
generate entangled states of photons among multiple cavities. In addition,
the approach presented in [80] for creating entangled states of photonic
qubits, which were encoded in both orthogonal polarization states and energy
eigenstates, was based on adiabatic passage techniques. In contrast, as
shown above, our present approach is based on resonant interaction.

\begin{center}
\bigskip

\textbf{IV. CONCLUSION}
\end{center}

We have presented a method to generate a GHZ state of $n$ photons in $n$
cavities coupled by a superconducting qutrit. By local operations on a qubit
(e.g., an atom etc.) placed in each cavity, the created GHZ states of
photons can be transferred to qubits for the storage for a long time. This
proposal is easy to be implemented in experiments since only resonant
qutrit-cavity interaction and resonant qutrit-pulse interaction are needed,
and no measurement is required. In addition, we have shown how to apply the
present method to create a GHZ state of $n$ photons in $n$ cavities via an
atom. We note that neither adjusting the atomic level spacings nor adjusting
the cavity mode frequency is needed during the entire operation and only one
atom is needed for the entanglement preparation of photons in multiple
cavities. In addition, our analysis shows that generating a GHZ state of
photons in up to four cavities by a coupler superconducting qutrit or a GHZ
state of photons in ten cavities via an atom is possible within the present
experimental technique. Finally, it should be mentioned that this proposal
is quite general, which can be applied to create a GHZ state of photons in
multiple cavities or resonators, when the coupler qutrit is a different
physical system, such as a quantum dot or an NV center.

\begin{center}
\textbf{ACKNOWLEDGMENTS}
\end{center}

C.P. Yang was supported in part by the National Natural Science Foundation
of China under Grant No. 11074062, the Zhejiang Natural Science Foundation
under Grant No. Y6100098, the Open Fund from the SKLPS of ECNU, and the
funds from Hangzhou Normal University. Q.P. Su was supported by the National
Natural Science Foundation of China under Grant No. 11147186. S. Han was
supported in part by DMEA.


\begin{references}
\bibitem{s49} For a phase qubit with the three levels considered, the $\left|
0\right\rangle \leftrightarrow \left| 2\right\rangle $ dipole matrix
element is much smaller than that of the $\left| 0\right\rangle \leftrightarrow
\left| 1\right\rangle $ and  $\left|
1\right\rangle \leftrightarrow \left| 2\right\rangle $ transitions. Thus, $\gamma _{20}^{-1}\gg \gamma _{10}^{-1},\gamma _{21}^{-1}.$
\end{references}
\end{document}